\documentclass[a4paper]{article}
\usepackage{color,float,epsfig,endfloat,changebar,newcent,textcomp,natbib}

\bibpunct{[}{]}{;}{}{,}{,}
\begin{document}

\title{Extended Recurrence Plot Analysis and its Application to ERP Data}
\author{Norbert Marwan$^{1,}$\thanks{email: marwan@agnld.uni-potsdam.de}, Anja Meinke$^2$\vspace*{6pt} \\
\small{$^1$ Nonlinear Dynamics Group, Institute of Physics,}\\ 
\small{University of Potsdam, Potsdam 14415, Germany}\\
\small{$^2$ Institute of Linguistics,}\\
\small{University of Potsdam, Potsdam 14415, Germany}}

\date{September 18, 2002}


\maketitle

\begin{abstract}
We present new measures of complexity and their application
to event related potential data. The new measures base
on structures of recurrence plots and makes the identification
of chaos-chaos transitions possible. The application of these 
measures to data from single-trials of the Oddball experiment 
can identify laminar states therein. This offers a new way of 
analyzing event-related activity on a single-trial basis.
\end{abstract}

\section{Introduction}
Neurons are known to be nonlinear devices because they become activated 
when their somatic membrane potential crosses a certain threshold 
\citep{kandelschwartz95}. This nonlinearity is one of the essentials 
in neural modelling which leads to the sigmoidal activation functions of 
neural networks \citep{amit89}. The activity of large formations of 
neurons is macroscopically measurable as the electroencephalogram 
(EEG) at the human scalp which results from a spatial integration of 
postsynaptic potentials \citep{nunez81}. However, it is an unsolved 
problem whether the EEG should be treated as a time series stemming 
from a linear or a nonlinear dynamical system. Applying nonlinear 
techniques of data analysis to EEG measurements has a long tradition. 
Most of these efforts have been done by computing the correlation 
dimension of spontaneous EEG \citep[e.\,g.][]{babloyantz85, 
rapp86, gallez91, lutzenberger92, pritchard92}. 
\citet{theiler92} applied 
the technique of surrogate data to correlation dimensions of EEG and 
reported that there is no evidence of low-dimensional chaos but 
of significance for nonlinearity in the data.

While correlation dimensions are only well defined for stationary 
time series generated by a low-dimensional dynamical system moving 
around an attractor, these measures fail in investigating event-related 
brain potentials  \citep[ERPs, ][]{sutton65} since they are nonstationary by definition. 
Traditionally, ERP waveforms are determined by computing an ensemble 
average over a collection of stimulus time locked EEG trials. 
This is based on the following assumptions:
(1) the presentation of stimuli of the same kind is followed by 
the same sequence of processing steps,
(2) these processing steps always lead to activation of the same 
brain structures,
(3) this activation always elicits the same pattern of 
electrophysiological activity, which can be measured at the scalp 
\citep{roesler82} and
(4) spontaneous activity is stationary and ergodic \citep{beimgraben00}.

By averaging the data-points time-locked to the stimulus 
presentation \linebreak (cf.~Oddball experiment) it is possible to filter out the signal (ERP) 
of the noise (spontaneous activity). In the next step 
the functional significance of a component is assessed. 
Antecedent conditions of the occurrence of a component 
and variables, which influence its parameters are defined. 
Now the commonalities of these factors are identified. The 
generalization of all empirically found influencing factors 
leads to a more abstract cognitive theory of the functional 
meaning of a event-related potential component and 
makes it usable for the validation of models of cognitive 
processes.

The disadvantage of the averaging method is the high number 
of trials needed to reduce the signal-to-noise-ratio
\citep{kutas84}. This is crucial for example for 
clinical studies,  for studies with children and for studies, 
in which repeating a task would influence the performance.
So it is desirable to find new ways of analyzing event-related 
activity on a single-trial basis. Applying nonlinear methods 
to electrophysiological data could be one way of dealing with 
this problem.

To compute dimensions of ERPs, \citet{molnar95} used the 
pointwise dimensions and reported a drop of the pointwise dimension 
as a function of time corresponding to the P300 component observed 
in the Oddball experiment. Recently, concepts of 
information theory have been introduced to 
analyse ERPs. On one hand this is the wavelet entropy of 
\citet{quiroga01} and on the other hand symbolic 
dynamics of EEG and ERP \citep{beimgraben00, 
frisch02, steuer02, schack02}.

A further promising approach is the recurrence quantification
analysis (RQA), which is based
on the quantification of the diagonal oriented structures
in recurrence plots \citep[RPs, ][]{webber94, zbilut92}.
The RQA was broadly applied in
a wide field of the analysis of physiological data 
\citep[e.\,g. ][]{casdagli97,faure98,thomasson2001,marwan2002herz}.
The important advantage of methods based on the quantification of RPs
is that the required data length can be relatively short.
However, the measures of the classical RQA are only able to
recognize transitions between periods and chaos and vice versa
\citep{trulla96}. In this work, we will use recently introduced
additional measures based on RPs in order to find
chaos-chaos transitions in physiological data. These
new measures use the vertical structures in the RP
and are able to identify laminar states \citep{marwan2002herz}.

In the first section we will give a short introduction
into RPs and their quantification analysis. In the next section
we will introduce the new measures and finally we will apply
them to event related potential data gained from the Oddball
experiment.

\section{Recurrence Plots and Their Quantification}

The method of recurrence plots (RP) was introduced to visualize the time
dependent behavior of the dynamics of systems,
which can be pictured as a trajectory in the phase space \citep{eckmann87}.
It represents the
recurrence of the $m$-dimensional phase space trajectory $\vec x_{i} \in 
\mathcal{R}^m$ 
($i=1, \ldots, N$, time discrete) to a certain state. The main
step of this visualization is the calculation of the $N \times N$-matrix
\begin{equation}
\mathbf{R}_{i,\,j} := \Theta(\varepsilon_i-\|\vec x_{i} - \vec x_{j}\|),
\quad \, i, j=1\dots N,
\end{equation}
where $\varepsilon_i$ is a state dependent cut-off distance, 
$\| \cdot \|$ is the norm of vectors, $\Theta$ is the Heaviside function
and $N$ is the number of states. The phase space vectors for 
one-dimensional time series $u_i$ from observations can be 
reconstructed with the Taken's time delay method
$\vec x_i=( u_i, u_{i+\tau}, \dots, u_{i+(m-1)\,\tau})$
with dimension $m$ and delay $\tau$ \citep{kantz97}.
The recurrence plot exhibits characteristic large-scale and small-scale
patterns which are caused by typical dynamical behavior
\citep{eckmann87, webber94}, e.\,g.~diagonals (similar local time evolution of
different parts of the trajectory) or horizontal and vertical black lines
(state does not change for some time). 

Zbilut and Webber have developed the recurrence quantification
analysis (RQA) to quantify an RP \citep{webber94,zbilut92}.
They defined measures using the recurrence point density and {\it diagonal}
structures in the recurrence plot, the recurrence rate $RR$ (density of
recurrence points), the determinism $DET$ (ratio of recurrence points
forming diagonal structures to all recurrence points), the maximal
length of diagonal structures $L_{max}$ (or their averaged length
$L$), the Shannon entropy $ENT$ of the distribution of the
diagonal lengths and the trend $TREND$ (paling in the RP). The computation
of these measures in shifted windows  along the
main diagonal of the RP enables one to find
characteristic excursions of the trajectory in the           
phase space of the considered systems.

Trulla et al.~have applied these measures in order to find
transitions in dynamical systems \citep{trulla96}. 
They have showed, that the RQA
is able to find transitions between chaos and order
(periodical states). But they could not find the chaos-chaos
transitions.

\section{Laminarity and Trapping Time}

We have recently introduced two additional measures which are based on the
{\it vertical} structures in the RP \citep{marwan2002herz}.
We define these measures analogous to the definition of
$DET$ and $L$, but we consider the
distribution $P(v)$ of the length of the
vertical structures in the RP.

First, the laminarity $LAM$
\begin{equation}
LAM:=\frac{\sum_{v=2}^{N}vP(v)}{\sum_{v=1}^{N}vP(v)},
\end{equation}
is the ratio of recurrence points forming vertical structures to all
recurrence points and represents the probability of 
occurrence of laminar states in the system, but it does not describe
the length of these laminar phases. It will decrease if the RP
consists of more single recurrence points than vertical
structures. 

Next, the trapping time $TT$
\begin{equation}
TT := \frac{\sum_{v=2}^{N} v P(v)} {\sum_{v=2}^{N} P(v)},
\end{equation}
is the averaged length of the vertical structures. The measure
$TT$ contains information about the amount and the
length of the laminar phases.

\begin{figure}[htbp]
\centering \epsfig{file=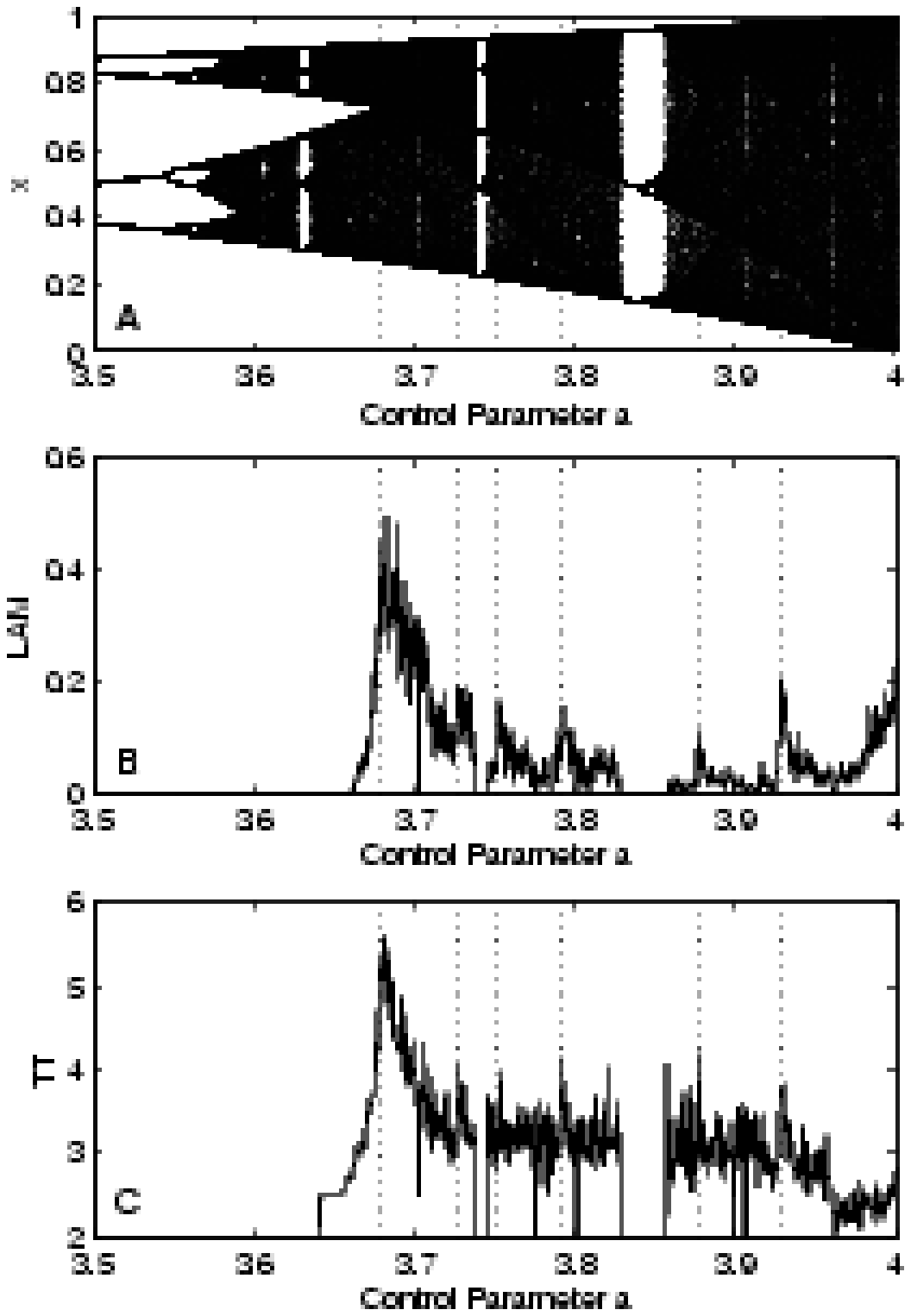, width=10cm}
\caption{Laminarity (B) and trapping time (C) of time series gained
from the logistic map for various control parameters (A). 
These measures reveal laminar and intermittent states.
The vertical dotted lines show a choosing
of points of band merging and laminar behaviour 
($a=3.678$, $3.727$, $3.752$, $3.791$, $3.877$, $3.927$).
The length of the data were $N=1000$ and the embedding 
parameters were $m=1$, $\tau=1$ and
$\varepsilon=0.1$.}\label{rqa_log}
\end{figure}

The difference between these measures and the traditional RQA
measures is their ability to find transitions between chaos and
chaos \citep{marwan2002herz}. For example, such transitions 
can be found in the logistic map $x_{n+1}=a\,x_{n}\left( 1-x_{n}\right)$
with increasing control parameter $a \in [0,4]$ and
$x_n \in [0,1] \subset \mathcal{R}$. For such trajectories $x(a)$ which contain
laminar states (e.\,g.~$a=3.678, 3.791, 3.927$),
$LAM$ and $TT$ show pronounced maxima (Fig.~\ref{rqa_log}).
The application of these measures to heart rate variability data, 
has shown, that they are able to detect and
quantify laminar phases before a life-threatening cardiac
arrhythmia and, thus, to enable a prediction of such an event
\citep{marwan2002herz}. These findings can be of importance for 
the therapy of malignant cardiac arrhythmias.

In the next section we will apply this
extended RQA to physiological data.

\section{Event Related Potentials}
\subsection{The Oddball experiment}

As mentioned in the Introduction, the Oddball experiment 
studies brain potentials during a stimulus presentation.

\begin{figure}[htbp]
\centering \epsfig{file=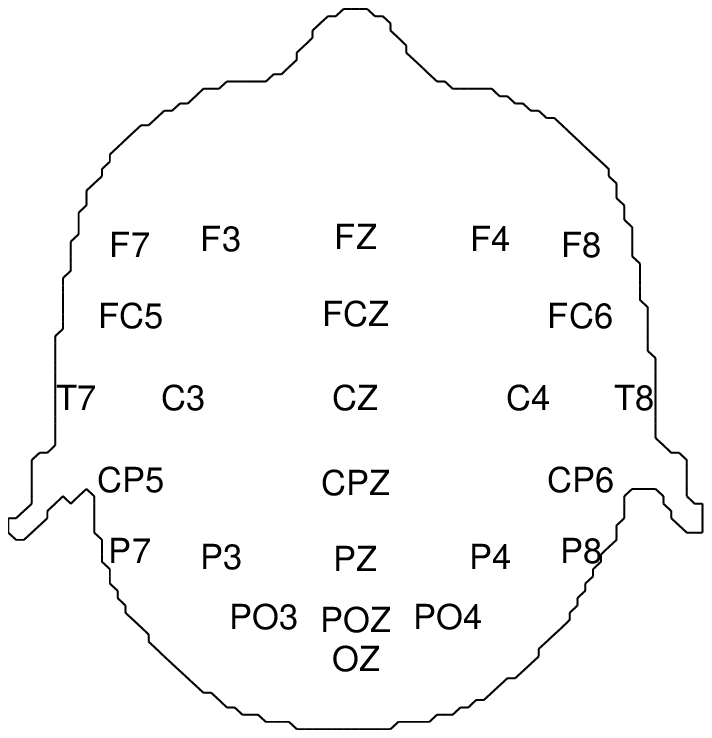, width=6cm} \\
\caption{Localization of the electrodes on the head.}\label{elc}
\end{figure}

The measurement of the EEG was done with 31 electrodes/ channels
(Tab.~\ref{elc_tab}). The first
25 electrodes were localized as shown in Figure~\ref{elc}; the others were
reference electrodes. The sample interval for the measurements was 4~ms.

\begin{table}
\centering
\caption{Notation of the electrodes and their numbering as it
is used in the figures (electrodes 26--31 are reference electrodes).}\label{elc_tab}
\begin{tabular}{rlrl}
\hline
\# &Electrode &\# &Electrode \\
\hline
1 &F7 &14&T8\\  2 &FC5&15&P7\\
3 &F3 &16&PZ\\  4 &FZ &17&P3\\
5 &F4 &18&CZ\\  6 &FC6&19&P4\\
7 &F8 &20&P8\\  8 &T7 &21&OZ\\
9 &CP5&22&POZ\\10&C3 &23&PO3\\
11&FCZ&24&CPZ\\12&C4 &25&PO4\\
13&CP6&&\\
\hline
\end{tabular}
\end{table}

Probands were seated in a dimly lit room in front of a monitor and
were instructed to count tones of high pitch. Each subject was tested
in nine blocks. The blocks varied in the probability of occurrence of
the higher tones from 10 to 90\,\%. Each block contained at least
30 target tones. Response was given in a three alternative choice 
(using cursor keys of the keyboard). During the test, the EEG was recorded.
The stimuli were computer-generated beeps of 100~ms length. Tones were
either high (1400~Hz) or low (1000~Hz). They were presented with an
interstimulus interval of 1000~ms.

\begin{figure}[htbp]
\centering \epsfig{file=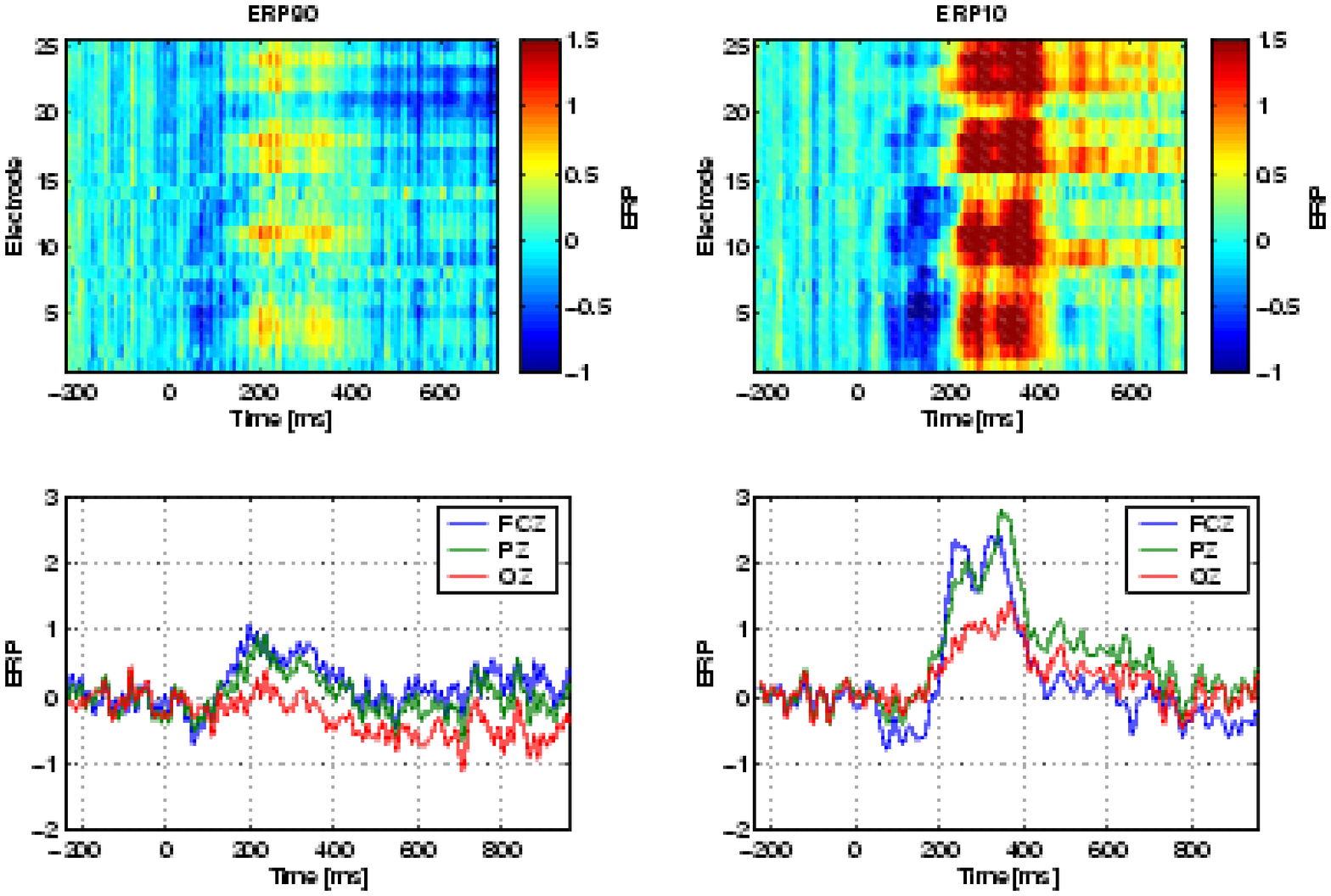, width=12cm}
\caption{Mean event related potentials 
for event frequencies of 90\,\% (left, 40 trials) and 
10\,\% (right, 31 trials). The N100 and
P300 components are well pronounced for the frequencies 
of 10\,\%. The lower plots show the ERP of selected electrodes.
The reference of the electrode numbers is given
in Table~\ref{elc_tab}.}\label{ekp_m}
\end{figure}

After computing event-related voltage averages for the experimental
manipulations (10\,\% up to 90\,\% target probability) one can observe a P300
ERP component whose amplitude is anti-correlated to the probability of the
stimuli (surprise ERP, Fig.~\ref{ekp_m}).

The P300 component of the ERP was the first potential discovered to vary in     
dependence on subject-internal factors like attention and expectation instead   
of physical characteristics \citep{sutton65}. The amplitude of the P300    
component is highly sensitive to novelty of an event and its relevance. So this 
component is assumed to reflect the updating of the environmental model of the  
information processing system \citep[context updating, ][]{donchin81,donchin88}.

\subsection{Data analysis}

Our focus will be directed to the ERP data of
two extreme event probabilities.
Henceforth, the time (measured in ms) is denoted as $t$,
the trial number as $i$ and the electrode as $e$ (the
allocation of the electrode numbers with their notion, see
Fig.~\ref{elc}).

\begin{figure}[htbp]
\centering \epsfig{file=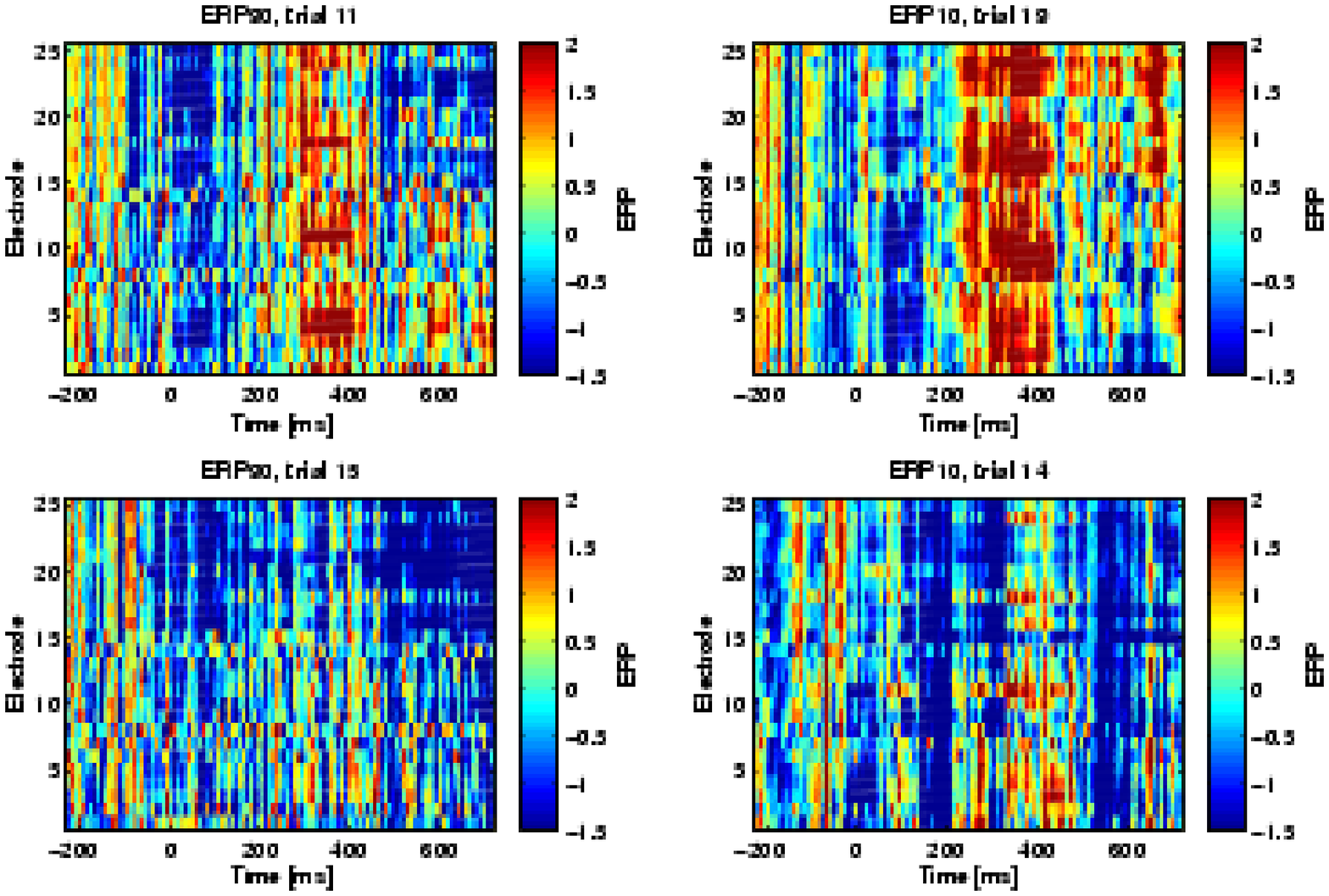, width=12cm}
\caption{Event related potentials for selected
trials of the event frequencies of 90\,\% (left) and 
10\,\% (right). Both, ERP10 and ERP90 of single trials
can be strongly or weakly pronounced, respectively, which
makes their discrimination difficult.
The reference of the electrode numbers are given
in Table~\ref{elc_tab}.}\label{ekp}
\end{figure}

The first set ERP90 contains 40 trials of ERP data for an
event frequency of 90~\%  and the second set ERP10
contains 31 trials
for an event frequency of 10~\%. Both
data sets can be rather well discriminated in the N100 and P300
components by the average over all trials (Fig.~\ref{ekp_m}).
As expected, both components have increased for lower event
probabilities (ERP10).
The maxima of the P300 are located around the central
and central-parietal electrodes. However, the single
trials do not obtain such a clear result. The P300 component 
is only well pronounced in 15 trials. When the single
trials are observed, then extreme values can also occur
in the ERP90 data and vanish in the ERP10 data (Fig.~\ref{ekp}).
We applied also a statistical variance-based T-test 
to the single trial ERP data. However,
this method could also not clearly distinguish the single trials.

\begin{figure}[htbp]
\centering \hspace*{.3cm} \epsfig{file=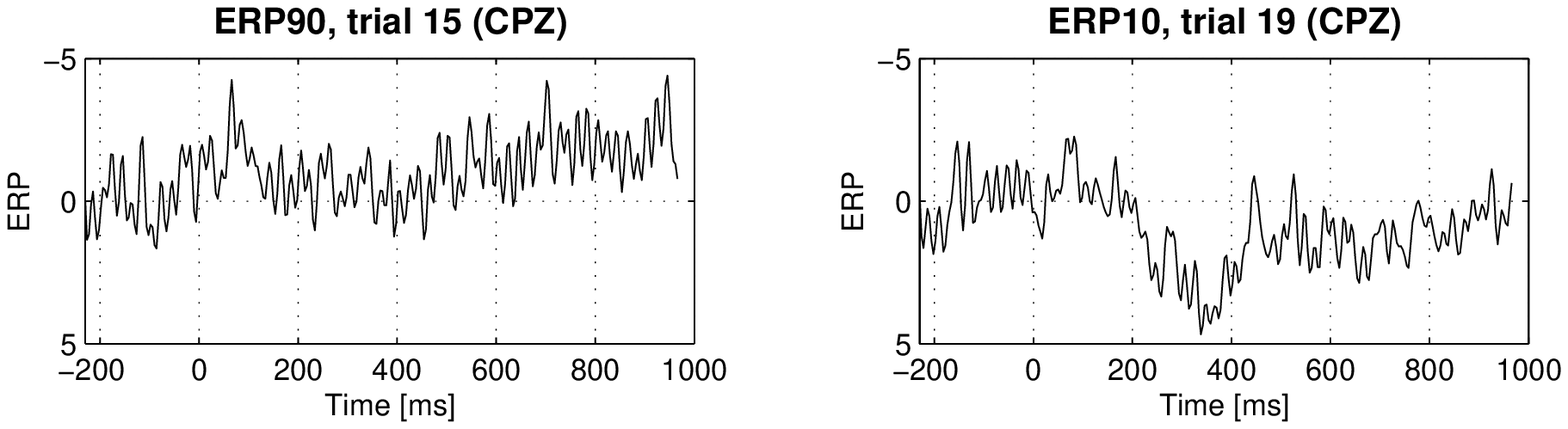, width=12cm}
\vspace*{.3cm}

\centering \epsfig{file=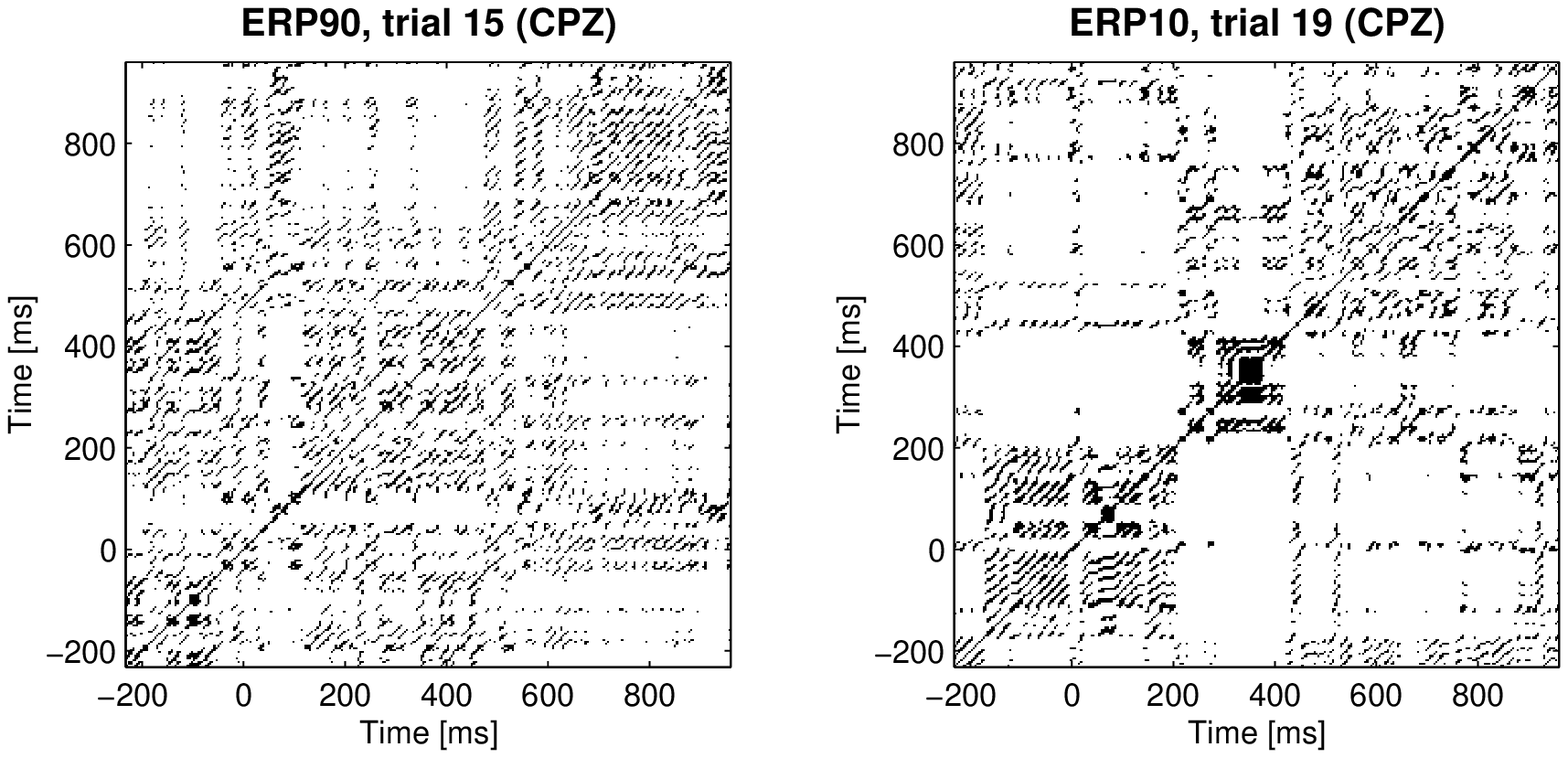, width=12cm} \\
\caption{ERP data for event frequencies of 90\,\% (upper left) and 
10\,\% (upper right), and their corresponding recurrence plots 
(lower plots). For the lower event frequency, more cluster of
recurrence points occur at 100ms and 300ms.}\label{rps}
\end{figure}

The recurrence quantification analysis (RQA)
is based on the structures obtained by recurrence
plots (RPs). The RPs were firstly computed for the
means of ERP90 and ERP10 over all trials and then
for the single trials. This
was done with the embedding parameters $m=3$,
$\tau=3$ and $\varepsilon=10\,\%$ (fixed amount of nearest neighbours).
The embedding parameters were estimated by using the
standard methods false nearest neighbours (dimension)
and mutual information (delay) \citep{kantz97}.
Due to the N100 and the P300 components in the data,
the RPs show varying structures changing in time (Fig.~\ref{rps}).
Diagonal structures and clusters of black points occur. The
nonstationarity of the data around the N100 and P300 causes
extended white bands along these times in the RPs. However,
the clustered black points around 300~ms occur in almost all
RPs of the ERP10 data set.

The RQA was computed from the RPs of ERP90 and ERP10 
for the single trials, in sliding windows over the RPs 
(which have the dimension $m=3$) with a length of
240~ms and with a shifting step of 8~ms. This window length
corresponds with a data length of 60 values.

\begin{figure}[htbp]
\centering \epsfig{file=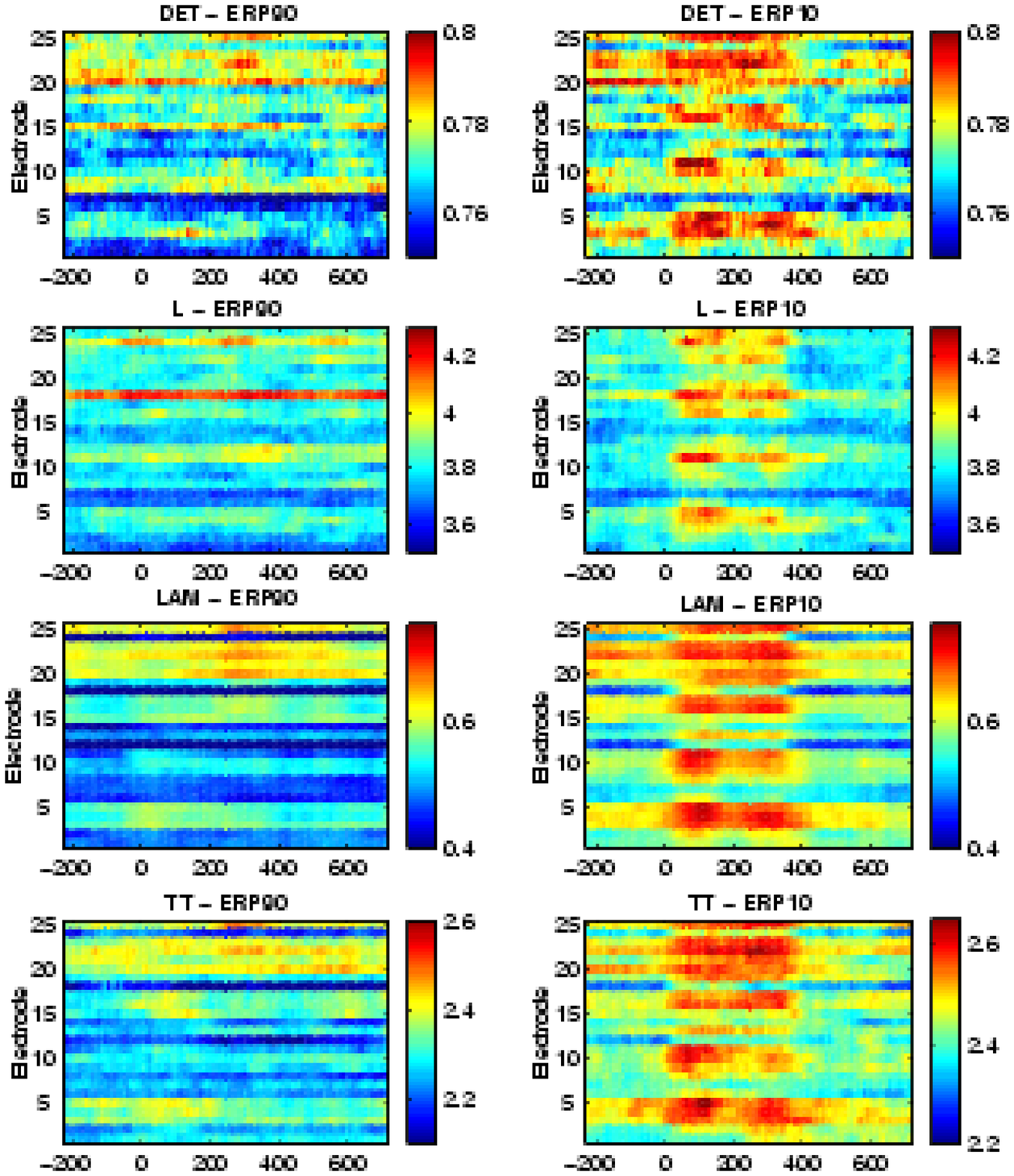, width=12cm} \\
\caption{Averaged RQA measures for the ERP data of both 
event frequencies (averaged over all trials). Whereas 
the measures do not reveal any transitions in the ERP90 data,
they clearly recognize the transitions for the ERP10 data.}\label{rqa_m}
\end{figure}

The mean of all RQA variables of ERP10 reveal
typical structures in the data (Fig.~\ref{rqa_m}, right column). They
indicate the transitions corresponding to the N100 and
P300 components around the central electrodes. 
The RQA variables for the ERP90 do not 
reveal these transitions (Fig.~\ref{rqa_m}, left column).
The onset of the increasing of the parameter is
about 120~ms before the event. This is due to the
windowed analysis of the RPs (240~ms windows). We have chosen
the middle of the RP window for the time, what results
in a 120~ms earlier onset of the RQA variables.

The four RQA variables are quite different, especially in their
amplitude.
For ERP10, $LAM$ and $TT$ are the best pronounced parameter and have
two distinct maxima at some electrodes; $DET$ and $L$ reveal
these maxima at these electrodes too, but are lesser
pronounced (Fig.~\ref{rqa_m}). These maxima occur
at the transition around 100~ms and 300~ms after the
event and occur at the electrodes F3, F4, FZ, C3, FCZ, PZ, POZ and PO3.
Differences between the various transitions found by these measures 
also occur in time and brain locations (electrodes). But, the study 
was not detailed enough in order to give reliable results.

\begin{figure}[htbp]
\centering \epsfig{file=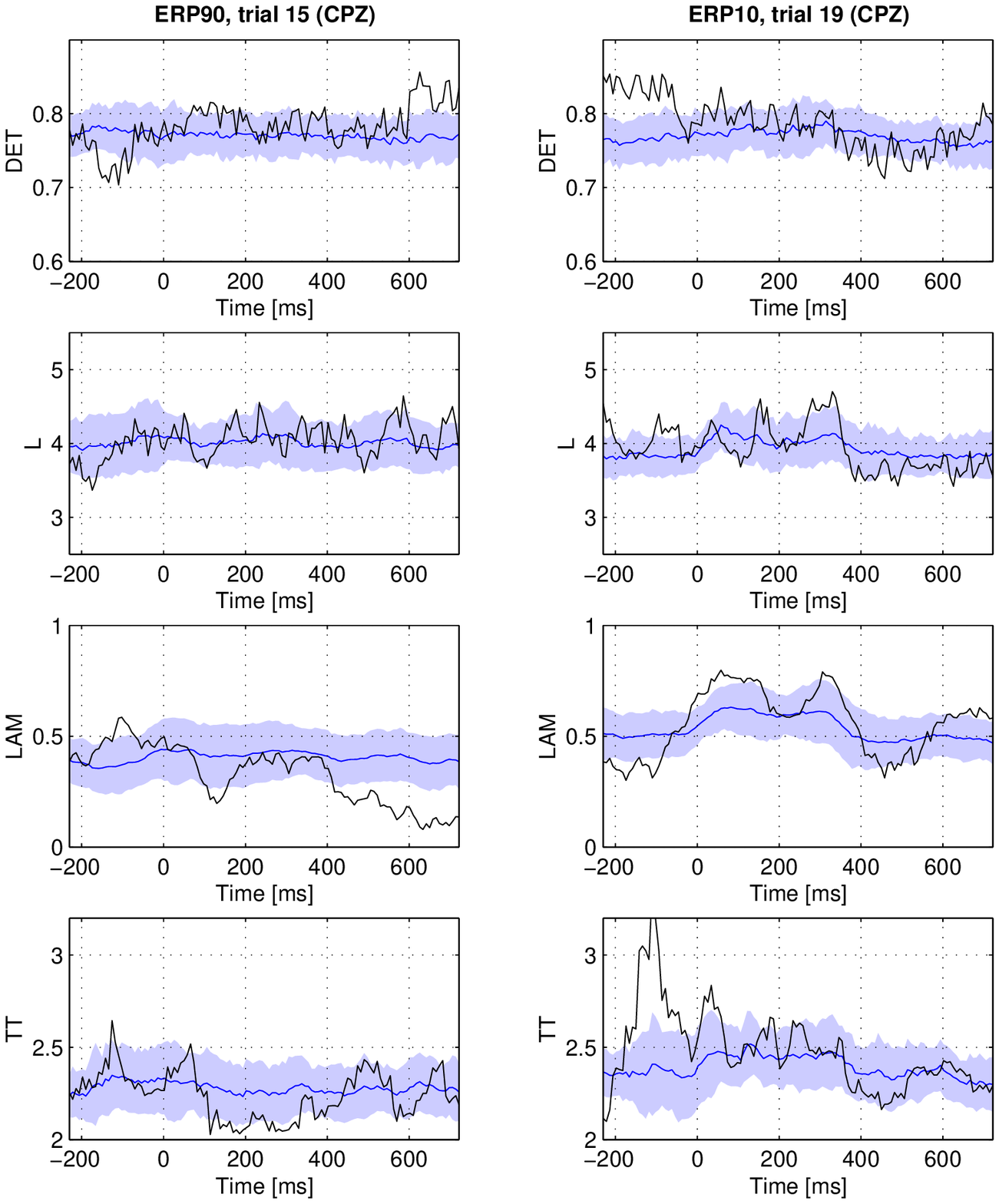, width=12cm} \\
\caption{RQA measures for selected single trials and the central-parietal
electrode (black). The trial-averaged RQA measures for the same electrode
is shown in blue (the light blue band marks the 95\,\% significance
interval).}\label{rqa_cpz}
\end{figure}

\begin{figure}[htbp]
\centering \epsfig{file=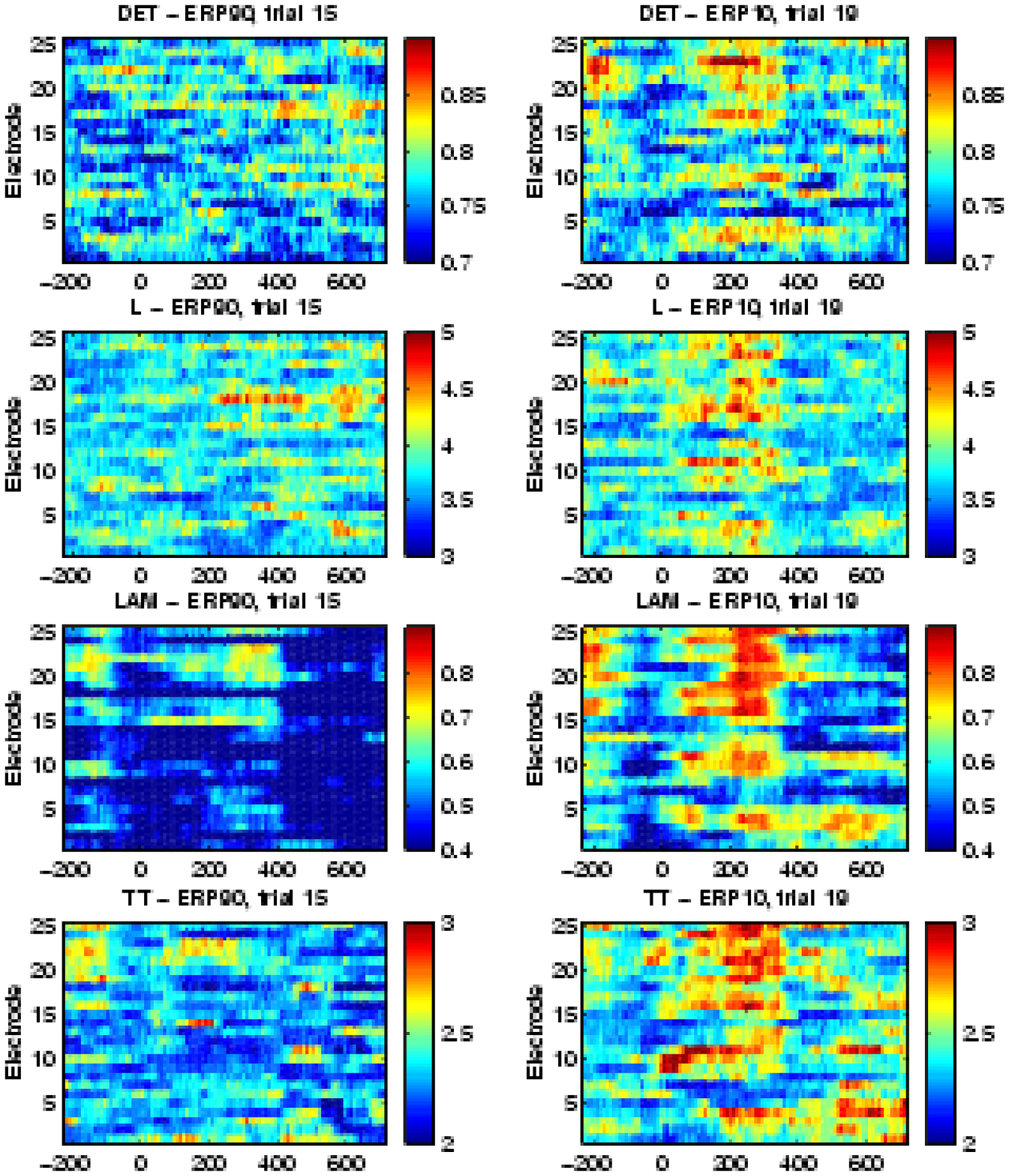, width=12cm} \\
\caption{RQA measures for the same trials as in Fig.~\ref{rqa_cpz},
but shown for all electrodes.}\label{rqa}
\end{figure}

The analysis of the single trials achieves
similar results (Figs.~\ref{rqa_cpz} and \ref{rqa}
show the results for selected trials).
The $LAM$ clearly found the
N100 and P300 components for ERP10 in 26 trials (of 31), 
but not for the ERP90 trials. The other measures have 
lesser maxima and, thus, are not suitable for such
recognition. 

This result indicates that our introduced measures of complexity
(especially $LAM$) are able to recognize transitions in brain potentials,
which are caused by e.\,g.~stimulative events. These transitions
can be found in the single trials, which is an improvement
to the classical method of averaging all observations.

\section{Summary}
We have applied an extended recurrence quantification analysis (RQA)
to physiological event related potential data (ERP).
The classical RQA consists of measures which are mainly based on
diagonal structures in the recurrence plots (RPs), e.\,g.~the
determinism ($DET$), which is the ratio of recurrence points
located on connected diagonal structures in the RP, and the
averaged diagonal line length ($L$). We have extended the
RQA with two recently introduced measures, the laminarity ($LAM$)
and the trapping time ($TT$). These measures are analogously
defined as $DET$ and $L$, but provided by
the vertical structures in recurrence plots. Whereas
the classical RQA  enables the identification of period-chaos
transitions, the new measures make the identification of 
chaos-chaos transitions and laminar states possible.

The classical method to study ERP data is to average them over
many trials. Our aim was to study the single trials in order
to find transitions in the data. 

The application of the
extended RQA to ERP data has discriminated the single trials with a distinct
P300 component due to a high surprise moment (less frequent
events) against such trials with a low surprise
moment (high frequent events). Considering the 
raw ERP10 data, the P300 component can only be found in the half
of all trials. Also a statistical variance test fails to distinguish 
cleary the trials. 
The $LAM$ is the most pronounced parameter
in this analysis. It measures the ratio of recurrence points
located on connected vertical structures in an RP. This
structures correspond with laminarity within the underlying
process. In the ERP data, the $LAM$ reveals transitions from less laminar
states to higher laminar states after the occurrence
of the event and a transition from higher laminar states
to less laminar states after about 400~ms. These transitions occur
around bounded brain areas (parietal to frontal along the central axis).
The comparable measures $DET$/ $LAM$ and $L$/ $TT$ are quite different in their
amplitude. There should also be differences in time and  
brain location of the found transitions. 

These results show that the measures based on vertical RP structures
make the identification of transitions possible, 
which are not found by the classical RQA measures. They 
indicate transitions in the brain processes
into laminar states due to the surprising moment of observed events. 

A future work will be concerned with the 
development of a statistical evaluation of these results.
Furthermore, this investigation has to be extended to 
ERP data gained from other frequent events and
a detailed study of the comparable measures $DET$/ $LAM$ and $L$/ $TT$
should give hints about the different transitions in the brain processes.

\section{Acknowledgments}
This work was partly supported by the priority programme SPP\,1097 
of the German Science Foundation (DFG).
We gratefully acknowledge the colleagues of the Nonlinear
Dynamics Group Potsdam for useful discussions and support of this 
work. 

\clearpage
\bibliographystyle{plainnat}
\bibliography{../mybibs,eeg}

\end{document}